\documentclass[prl,twocolumn,showpacs,preprintnumbers,
amsmath,amssymb,superscriptaddress]{revtex4}

\usepackage{graphicx}
\usepackage{bm}

\newcommand{\beq}{\begin{equation}}
\newcommand{\eeq}{\end{equation}}
\newcommand{\beqa}{\begin{eqnarray}}
\newcommand{\eeqa}{\end{eqnarray}}
\newcommand{\w}{\omega}

\newcommand{\raiz}{\mbox{$\textstyle \frac{1}{\sqrt{2}}$}}
\newcommand{\ket}[1]{\left| #1 \right\rangle}

\newcommand{\upket}{\ket{\uparrow}}
\newcommand{\downket}{\ket{\downarrow}}

\begin{document}

\preprint{LRNB1102}
\title{Resonant Transfer of Excitons and  Quantum Computation}
\author{Brendon~W.~Lovett}
\affiliation{Department of Materials, Oxford University, Oxford OX1 3PH, United Kingdom}
\author{John~H.~Reina}
\altaffiliation[On leave from ]{CIF, A.A. 4948, Bogot\'a, Colombia.}
\affiliation{Department of Materials, Oxford University, Oxford OX1 3PH, United Kingdom}
\affiliation{Clarendon Laboratory, Department of Physics, Oxford University, Oxford OX1 3PU, United Kingdom}
\author{Ahsan~Nazir}
\affiliation{Department of Materials, Oxford University, Oxford OX1 3PH, United Kingdom}
\author{Beeneet Kothari}
\affiliation{Department of Materials, Oxford University, Oxford OX1 3PH, United Kingdom}
\author{G.~Andrew~D.~Briggs}
\affiliation{Department of Materials, Oxford University, Oxford OX1 3PH, United Kingdom}
\date{\today}

\begin{abstract}
Resonant energy transfer mechanisms have been observed in the sensitized
luminescence of solids, in quantum dots and in molecular nanostructures, and they also play a central role in light
harvesting processes in photosynthetic organisms.
We demonstrate that such mechanisms, together with the exciton-exciton binding energy shift typical of these
nanostructures, can be used to perform universal quantum logic.
In particular, we show how to generate controlled exciton entanglement and identify two different regimes of quantum
behaviour.
\end{abstract}


\maketitle

The F\"orster energy transfer was first studied in the context of the sensitized luminescence of
solids~\cite{forster59,dexter53}, in which an excited sensitizer atom can transfer its excitation to a neighbouring acceptor
atom, via an intermediate virtual photon.
This mechanism is also responsible for photosynthetic energy processes  in antenna complexes, biosystems (BSs) that harvest
sunlight~\cite{hu02}.
More recently, interest has focussed on such a transfer in quantum dot (QD) nanostructures~\cite{crooker02}
and within molecular systems (MSs)~\cite{hettich02}. In this Letter we give a general scheme for quantum computation
that can be implemented in different nanostructures~(NSs) by exploiting the F\"orster and exciton-exciton interactions.
Thus, methods for controlled generation of exciton entanglement that use both diagonal and off-diagonal interactions
are given.

Consider two coupled generic NSs (scalability will be addressed later). We assume that the excitations of each NS
are charge neutral (i.e., of an excitonic nature) and that they can be produced by optical means~\cite{gammon01}.
We also assume that tunnelling processes between them may be neglected,
but that there is a strong exciton-exciton electromagnetic coupling. Our two-level---{\it qubit} system  is represented
in each NS
by a single low-lying exciton (qubit state $\ket{1}$) and the ground state (qubit state $\ket{0}$).
Then the interaction Hamiltonian can be written
in the computational basis ($\{|00\rangle, |01\rangle,|10\rangle,|11\rangle\}$, with the first digit referring
to NS I and the second to NS II) as follows:
\beqa
\widehat{H}=
\left(
\begin{array}{cccc}
\w_0 & 0 & 0 &0 \\
0 & \w_0+\w_2 & V_{\rm F} & 0 \\
0 & V_{\rm F} &  \w_0+\w_1 &0 \\
0 & 0 & 0 &  \w_0+\w_1+\w_2+V_{\rm XX}
\end{array}\right)
\label{eq:Hdot}
\eeqa
where the diagonal interaction $V_{\rm XX}$ is the direct Coulomb binding energy
between the two excitons, one located on each NS, and the off-diagonal $V_{\rm F}$ denotes the Coulomb exchange
(F\"orster) interaction which induces the
transfer of an exciton from one NS to the other. These are the only Coulomb interaction terms which act between the
qubits. $\w_0$ denotes the ground state energy and we define $\Delta_0 \equiv \w_1-\w_2$ as the difference between the exciton
creation energy for NS I ($\w_1$) and that for NS II ($\w_2$).
Energy contributions due to spin singlet-triplet splittings do not significantly affect the
present gating scheme and such effects are dealt with elsewhere~\cite{lovett02}.

The eigenenergies and eigenstates of the interacting  qubit system are
$E_{00}=\w_0$,
$E_{01}=\w_0+\w_1-\frac{\Delta_0}{2}(1-A)$,
$E_{10}=\w_0+\w_1-\frac{\Delta_0}{2}(1+A),
E_{11}=\w_0+\w_1+\w_2+V_{\rm XX}$;
and
$\ket{\Psi_{00}}=\ket{00}$,
$\ket{\Psi_{01}}=c_1\ket{10}+c_2\ket{01}$,
$\ket{\Psi_{10}}=c_2\ket{10}-c_1\ket{01}$,
$\ket{\Psi_{11}} = \ket{11}$,
where $A=\sqrt{1+4(V_{\rm F}/\Delta_0)^2}$, $c_1 = \sqrt{(A+1)/2A}$
($\approx V_{\rm F}/\Delta_0$ for $V_{\rm F}/\Delta_0\ll 1$)
and $c_2 = \sqrt{(A-1)/2A}$.
The eigenenergies in the absence and presence of interqubit interactions are displayed in Figs.~\ref{QC}(a) and (b)
respectively. Fig.~\ref{QC}(c) shows $c_1$ and $c_2$ as a function of the ratio $V_{\rm F}/\Delta_0$.

Single qubit operations can be achieved by inducing
Rabi oscillations in the excitonic system~(e.g., see Ref.~\cite{kama01}).
The $V_{\rm XX}$ and $V_{\rm F}$
interactions lead to two regimes for achieving quantum entanglement. First, 
if the ratio
$V_{\rm F}/\Delta_0 \gg 1$, then after selectively
exciting NS I  and creating $\ket{10}$, the system will naturally evolve into one of the maximally entangled Bell states
$\raiz (\ket{10} \pm i\ket{01})$ and this evolution could then be stopped by suppressing the F\"orster coupling (see later).
Second, if $V_{\rm F}/\Delta_0\ll 1$, the computational basis states are essentially the eigenstates of the system,
and we can exploit the $V_{\rm XX}$ term, together with appropriately tuned laser pulses, to perform two-qubit logic
gates and hence to
generate controlled entanglement. From Fig.~\ref{QC}(b) we can see how to implement the $\textsc{cnot}$ gate;
e.g.,  $\textsc{cnot}_{12}(\ket{1}\ket{0} \mapsto \ket{1}\ket{1})$ can be achieved by illuminating the qubit system in the
state $\ket{10}$ with a $\pi$-pulse of energy  $\epsilon_{12}$. If we start in the ground state and
first apply a $\pi/2$ or $3\pi/2$ pulse at energy
$\w_1$, we create the states $\raiz (\ket{00}\pm\ket{10})$; if we now apply a $\pi_{\epsilon_{12}}$ pulse,
we generate  the maximally entangled states $\raiz (\ket{00}\pm\ket{11})$.

In order to illustrate our ideas, we shall first calculate the strength of the interactions in
a coupled QD molecule (in a previous study of QD excitons for quantum computing (QC)~\cite{biolatti02}, the off-diagonal interaction terms were neglected).
We start with the direct Coulomb interaction, $V_{\rm XX}$. Different dot geometries (e.g. spherical, pyramidal, cuboidal)
can be used to implement the scheme. In this Letter we assume that the dots are square-based cuboids and that the potential energy $V$ of both electrons and holes
increases abruptly at the
cube boundaries where the semiconductor bandgap changes, and  $V=0$ inside the cubes (see Fig.~\ref{dotspic}(a)).
This type of potential has the advantage of
both a well defined dot size in all three dimensions, and of
bound and unbound solutions in each direction (this is in contrast with the
parabolic potential considered in Ref.~[\onlinecite{biolatti02}]).
The wavefunctions for single particles may be expressed in the envelope
function approximation as
$\psi_p({\bf r}) = \phi_p({\bf r}) U_p({\bf r})$~\cite{harrison00},
where $\phi_p({\bf r})$ is an envelope function describing the changing wavefunction amplitude of confined states
for particle type $p$ over the dot region, and $U_p({\bf r})$ is the Bloch function which has the periodicity of the atomic
lattice. In the effective mass approximation, the
envelope functions are solutions of the single particle Schr\"odinger equation
for the potential $V_p$, and for the
effective mass  $m^\star_p$ of particle $p$ (see Fig.~\ref{dotspic}(a)).
These solutions may be obtained by expanding the Hamiltonian in a
set of envelope basis functions of the form $\Omega({\bf r}) = \xi_x(x)\xi_y(y)\xi_z(z)$,
(the $\xi_i(i)$ are the solutions
of a one-dimensional square well potential) and then diagonalizing~\cite{harrison00}.

We now combine the ground states of electrons and holes into a Slater determinant type wavefunction
which represents one exciton on each dot:
\beq
\Psi_{\rm XX} = \cal{A}[\psi_{\rm e}^{\rm I}({\bf r_1})\psi_{\rm h}^{\rm I}({\bf r_2})\psi_{\rm e}^{\rm II}({\bf r_3 - R})
\psi_{\rm h}^{\rm II}({\bf r_4 -  R})] \ ,
\label{XXwf}
\eeq
where $\cal{A}$ indicates that the wavefunction has overall antisymmetry,
${\bf R}$ is the vector connecting the two dot
centres, ${\bf r_1}$ and ${\bf r_3}$ represent the position vectors of electrons relative to the centres of
dot I and dot II respectively and ${\bf r_2}$ and ${\bf r_4}$ are the equivalent vectors for holes.
The expansion of the associated Coulomb operator $\hat{V}_{\rm XX}$ in a Taylor series about ${\bf R}$ to
lowest non-zero order yields
\beq
\hat{V}_{\rm XX} = \frac{k}{\epsilon_rR^3}\left\{{\bf p_{\rm I}}\cdot{\bf p_{\rm II}} -
\frac{3}{R^2}({\bf p_{\rm I}}\cdot{\bf R})({\bf p_{\rm II}}\cdot{\bf R})\right\} \ ,
\label{XXdipole}
\eeq
where $k\equiv\frac{1}{4\pi\epsilon_0}$, $\epsilon_r$ denotes the dielectric constant of the medium
($\epsilon_r=10$ throughout our discussion), ${\bf p_{\rm I}} = e({\bf r_1} - {\bf r_2})$, and ${\bf p_{\rm II}} = e({\bf r_3} - {\bf r_4})$ are the overall dipole
moments on dot I and II respectively.
To evaluate the matrix element $\langle \Psi_{\rm XX}|\hat{V}_{\rm XX}|\Psi_{\rm XX}\rangle$,
${\bf p_{\rm I}}$ and ${\bf p_{\rm II}}$ in Eq.~(\ref{XXdipole}) are replaced by their expectation values for the wavefunction,
Eq.~(\ref{XXwf}). This procedure gives rise to
a direct term and exchange terms. The exchange terms are zero in the absence
of wavefunction overlap between dots.
The direct term is obtained through the use of the envelope function
approximation which leads to the expectation value
$
\langle {\bf r_1} \rangle = \int \phi_{\rm e}^{\rm I\star} ({\bf r_1}){\bf r_1}\phi_{\rm e}^{\rm I} ({\bf r_1}) d{\bf r_1}
$,
and similar expressions hold for the other position expectation values.
For a cuboidal dot, where
the electron and hole wavefunctions have a definite parity about the dot centre,
Eq.~(\ref{XXdipole}) implies that
the exciton-exciton coupling is zero. However, when this symmetry is broken,
the electron and hole localize in different parts of the dot and the dipole moment is non zero.
This occurs, for instance, in pyramidal dots~\cite{pryor97} or when an electric field is
present~\cite{biolatti02,derinaldis02}.

We have simulated the effect of applying an electric field (along $x$) in our model
and the results are displayed in Fig.~\ref{dotspic}(b), where the exciton dipole moment
$p_i$ is plotted as a function of the dot size and the field strength $E$. Two cases are considered:
the first is that of a cubic dot ($a=h/2$) and the second is that of a flat cuboidal dot ($a=5h$).
$V_{\rm XX}$  is obtained by using the size of
$p_i$ for each dot, and Fig.~\ref{dotspic}(c) shows $V_{\rm XX}$ (normalized by  $R^{-3}$) calculated for both geometries,
with $a=b$. It is interesting that the size of the induced dipole depends strongly only on the length, $a$, of the
dot {\it in the direction of the applied field}, and at large field is given approximately by $p_i \approx ea$; this limit
should be valid for any NS (such as the BSs and MSs described later)
with a well defined lengthscale, and not just for QDs.
The cuboidal structure is more typical of Stranski-Krastanow self assembled dots, where typically $R=5$~nm, $a=10$~nm,
and $b=8$~nm, and $h_1=h_2=2$~nm.
In a field of 100~kV/cm, these parameters give $V_{\rm XX} = 120$ meV, which
would result in a  lower time limit for the gate operation
of around 10~fs. This is relatively short; decoherence times on the order of nanoseconds have been observed for
uncoupled dots~\cite{nsdeco}.

Let us now consider the F\"orster (off-diagonal) coupling $V_{\rm F}$ in QDs, which
may be expressed as a matrix element of the Coulomb operator between excitons located on each of the two dots.
By Taylor-expanding this expression around  ${\bf R}$ to lowest non-zero order~\cite{dexter53},
we obtain
$
V_{\rm F} = \frac{ke^2}{\epsilon_r R^3}\left(\langle {\bf r}_{\rm I} \rangle \cdot \langle {\bf r}{\rm_{II}} \rangle
-
\frac{3}{R^2} (\langle {\bf r}_{\rm I} \rangle \cdot {\bf R})(\langle {\bf r}_{\rm II} \rangle \cdot {\bf R})\right)
$. Here $e\langle {\bf r}_{i} \rangle$ ($i=\text{I, II}$) represents the  matrix element of the position operator between an
electron and a hole state on dot $i$.
The expression is therefore equivalent to the interaction of two point dipoles, one situated on each dot.
However, its magnitude is quite different from $V_{\rm XX}$.
By again employing the envelope function
approximation, we can rewrite the above equation as
\begin{equation}
V_{\rm F} = \frac{ke^2}{ \epsilon_r R^3} O_{\rm I} O_{\rm II}
\left(\langle{\bf r}_{\rm I}^{\rm a} \rangle\cdot\langle {\bf r}_{\rm II}^{\rm a} \rangle
-\frac{3}{R^2} (\langle {\bf r}_{\rm I}^{\rm a}
\rangle \cdot {\bf R}) (\langle {\bf r}_{\rm II}^{\rm a} \rangle \cdot {\bf R})\right)  ,
\label{envdipole}
\end{equation}
where $\langle {\bf r}_{i}^{\rm  a} \rangle = \int_{cell} U_e^i({\bf r}) {\bf r} U_h^i ({\bf r}) d{\bf r}$
is the interband position matrix element for the atomic part of the wavefunction for dot $i$, and
$O_{i} = \int \phi_{\rm e}^i ({\bf r}) \phi_{\rm h}^i ({\bf r}) d{\bf r}$ is the
overlap of electron and hole envelope functions on dot $i$.
Thus, the effects of the QD size and shape (which determine the overlap integrals) are separated from the effects of the
material composition of the dot (which determine the interband position operator). For other, smaller, NSs, the
envelope function approximation does not apply. However, in these cases the above analysis is still valid
if the overlap integrals are set to unity, and if the NS's entire wavefunction is represented by the atomic basis
part of the general wavefunctions described above.

We now introduce a simple Kronig-Penney model to describe the atomic basis, and we assume that the infinite Kronig-Penney
quantum boxes have a width of $2x$. This gives $\langle{\bf r}_{i}^{\rm  a}\rangle = 32x/9\pi^2$, or $V_{\rm F}\propto x^2$,
and so we plot
$V_{\rm F}/x^2$ as a function of $R$ in Fig.~\ref{newforster}(a). The simple Kronig-Penney model shows how the size of the
F\"orster transfer depends upon the physical size of the atomic part of the wavefunction.
$\langle{\bf r}^{\rm a}\rangle$ is a widely measured quantity since it determines the strength of dipole allowed
transitions in optical spectra. In CdSe QDs it can be in the range of 0.9 to 5.2~$e$\AA~\cite{crooker02},
in atomic systems it can also be several $e$\AA~\cite{NISTsite} and in BSs and MSs has recently been observed to be
about 1.7~$e$\AA~\cite{hettich02}.
The solid line in Fig.~\ref{newforster}(a) represents the case when the overlap integrals are set to unity;
the symbols represent
simulations which take higher order terms into account and have $a=b=2$~nm~\cite{lovett02}.
Fig.~\ref{newforster}(b) shows how $V_{\rm F}$ varies as a function of dot size and confinement potential
through the $O_i$'s of Eq.~(\ref{envdipole}).
The envelope functions are as described in the previous section, except that
no electric field term has been included in the envelope Hamiltonian. Thus the values of $V_{\rm F}$ should be
regarded as an upper limit for the cuboid dot model: applying a field will serve to reduce $V_{\rm F}$
since it decreases the $O_i$'s; hence it could be used to tune $V_{\rm F}$.
The overlap is enhanced when there is a
larger confinement potential and for larger dots, since in these cases the shape of the wavefunction is less
sensitive to the effective mass of the particle. As an illustration of the use of these curves, let us assume
that we have a dot system in which, as before, $R=5$~nm, $a=10$~nm, $b=8$~nm, and $h_1=h_2=2$~nm. Furthermore, let us
take the measured dipole value for CdSe dots of 0.9 to 5.2~$e$\AA~\cite{crooker02}. In this case, the F\"orster strength is
between $0.02$ and $0.6$~meV, which if $\Delta_0 = 0$ would correspond to an {\it on resonance} energy transfer rate of
between $206.8$  and $6.9$~ps. This is short enough to be useful for QIP: decoherence times
as long as a few ns~\cite{nsdeco} have been observed in QDs. In MSs or BSs, the interacting units
can be as close together as 1~nm; using this and taking a typical molecular or biomolecular dipole value of about
1.7~$e$\AA~\cite{hettich02, crooker02}, we obtain an interaction strength of 8.3~meV (or a transfer time of 497~fs).
Furthermore,  $V_{\rm F}$ must certainly be controlled if the alternative scheme  using $V_{\rm XX}$ is to be implemented (and therefore cannot be neglected as in Ref.~\cite{biolatti02}).

By also using the model outlined above to calculate $\Delta_0$ in QDs (shown in Fig.~\ref{Delta}(a) as a
function of dot size ratio for quantum cubes)
we can compute the size of the $c_1$ component of the $\ket{\Psi_{10}}$ and $\ket{\Psi_{01}}$ states.
This is shown in Fig.~\ref{Delta}(b), where the dependence on atomic dipole and
dot separation are incorporated by multiplying $c_1$ by $R^{3}x^{-2}$.
It can be seen there that a range of $c_1$ values can be obtained by choosing dots with appropriate values of
$x$, $R$ and $a/b$. Dots with large $x$ ($>1$~nm say), small $R$ ($<3$~nm say) and $a/b \sim 1$ give a larger $c_1$,
and it is then more appropriate to use the F\"orster interaction itself to create entangled states.
On the other hand, dots with smaller $x$,
larger $R$ or a large mismatch in dot size would be more suited to the scheme which uses the
$V_{\rm XX}$ for QC and  entanglement generation. The fidelity of a typical operation (e.g. $\ket{11}
\mapsto \ket{10}$) in this case is equal to $1-c_1^2$---and so one must be careful when using the biexciton scheme
to use the available parameter space and make sure that the F\"orster transfer is suppressed to the desired accuracy.
There are other sources of decoherence in this case (e.g. the interaction with optical and acoustic
phonons~\cite{nsdeco,lovett02}) which will reduce the value of the fidelity to below $1-c_1^2$.
To minimize the effects of such decoherence channels, it is important to maximize $V_{\rm XX}$,
since this leads to an improved transition discrimination and so to a faster gating time.
This can be done by applying an electric field and choosing an appropriate dot shape, size
and separation (as described earlier).
It is then necessary to minimize the basis state mixing for the chosen parameters by selecting a
suitable dot size ratio and material composition.


Single shot qubit state measurement in QDs could be performed by using resonant fluorescent shelving
techniques~\cite{blatt99}.
The QD state measurement can also be achieved by means of projecting onto the computational basis and measuring the
final register state by exploiting ultrafast near-field
optical spectroscopy and microscopy~\cite{gammon01}: these allow one  to address, to excite and to probe
the QD excitonic states with spectral and spatial selectivity. In addition,
the qubit  register density matrix can be reconstructed by measuring
the QD photon correlations via standard quantum state tomography techniques~\cite{bill02s}.
Scalability of the scheme given here could also be possible by adopting a globally addressed qubit
strategy~\cite{benjamin00} on a stack of self-organized QDs~\cite{xie95}.

Light-harvesting antenna complexes~\cite{hu02} or arrays of strongly interacting individual molecules~\cite{hettich02} could provide an excellent system in which the F\"orster interaction could be
used for QIP tasks. They are generally very uniform structures,
and we may compare them to QDs by setting $a/b \sim 1$, or
$V_{\rm F}/\Delta_0\gg 1$. Then the one-exciton eigenstates of a two qubit system with a F\"orster coupling naturally
allows the generation of the states $\raiz (\ket{01}\pm\ket{10})$, which, apart from their applications to quantum
protocols, can be particularly useful in the fight against decoherence. Spectroscopic, line-narrowing techniques
(e.g., hole burning and site-selective fluorescence), infrared and Raman experimental studies reveal that the main
decoherence mechanisms in  the antenna complexes arise from energetic disorder, electron-phonon coupling, and temperature
effects~\cite{hu02}.   In this scenario, the excitations couple to an environment that typically possesses a much larger
coherence length than the biomolecular units (BChl's) spacing. For example, the BChl's in the antenna complex LH2,
which we consider as potential qubits, are spaced by as little as 1~nm, and hence so-called collective decoherence is
expected to apply. In this case, provided that the logical qubit encoding $\downket_i\equiv\ket{01}_{jk}$,
$\upket_i\equiv\ket{10}_{jk}$ that uses two physical (exciton) qubits can be realized in the BChl's system,
arbitrary superpositions of  logical
qubits such as $(\alpha_i\downket_i+\beta_i\upket_i)^{\otimes N}$, $i=1,\ldots, N$, $\alpha_i$, $\beta_i\in\mathbb{C}$,
are immune to dephasing noise (described by a $\sigma_z$ operator~\cite{lidar98}), and single qubit manipulations can be
carried out on the timescale of the
F\"orster coupling (which as we have seen can be as short as 497~fs). Two-qubit logic gates can also be implemented
within  a decoherence-free subspace by using the above encoding, thus completing a universal set of gates~\cite{lidar98}.
Initialization of the system requires the pairing of the physical qubits to the logical `ground' state
$\ket{\downarrow}_i^{\otimes N}$, and  readout is to be accomplished by  identifying on which of the two structures the
exciton is. Furthermore, rings of BChl's appear side by side in naturally occuring antenna complexes
and also display energy selectivity---smaller rings tend to
have higher energy transitions~\cite{hu02}. Thus, following a scheme as above,  it may be possible to scale up such biological qubits in a natural way and construct a robust energy selective scheme for quantum computation.

In conclusion,
we have provided a general scheme for quantum computation and quantum entanglement generation that can be implemented in different NSs by exploiting the F\"orster and exciton-exciton interactions. In particular, we have shown how such interactions can be manipulated in molecular, biomolecular and QD nanostructures in order to produce an accurate degree of control for quantum logic.

BWL thanks St Anne's College for a Junior Research Fellowship; JHR and AN are supported by EPSRC (JHR as part of the
Foresight LINK Award {\it Nanoelectronics at the Quantum Edge}).
We thank T.~Spiller, W.~Munro, S.~Benjamin and R.~Taylor for stimulating discussions.

\begin{figure}[h]
\caption{A schematic diagram of the proposed quantum logic gate scenario:
energy levels in  (a)  the absence and (b) the presence of the interactions $V_{\rm XX}$ and $V_{\rm F}$ for nanostructures
of different  excitation frequency. $\epsilon_{12}\equiv \w_2+V_{\rm XX}-\delta$,
$\epsilon_{21}\equiv \w_1+V_{\rm XX}+\delta$, $\delta\equiv V_{\rm F}^2/\Delta_0$; (c) dependence of the eigenstate
coefficients $c_i$ as a function of $V_{\rm F}/\Delta_0$. The inset shows the  eigenenergies $E_{01}$, and $E_{10}$ of
plot (b) for $\w_1/\Delta_0\equiv 20$.
}
\label{QC}
\end{figure}

\begin{figure}[h]
\caption{(a)~Schematic diagram of the cuboidal dot model. The potential inside the cuboids is set to zero, that outside
is determined by the band offsets of the conduction and valence bands within the heterostructure.
(b)~Exciton dipole moment as a function of the dot size and applied electric field for two dot shapes. The dot parameters $m_e=0.06$, $m_h=0.6$, $V_e=V_h=500$~meV. (c)~Exciton-exciton binding energy for $a=b$
and sequence of dot shapes, size and field strength as in (b).}
\label{dotspic}
\end{figure}

\begin{figure}[h]
\caption{(a) Dependence of the F\"orster interaction strength on the interdot separation.
The solid line represents the case where  $O_i = 1$ in the dipole-dipole approximation and the
atomic dipole operator is given by $32x/9\pi^2$.
The symbols result from a full numerical simulation for $a=b=2$~nm, and the dotted lines are the dipole-dipole
predictions in these cases.
(b) The $O_i$ factor appearing in Eq.~(\ref{envdipole}) as a function of dot size and confinement potential.
$m_e=0.06\,m_0$, $m_h=0.6\,m_0$. A lower cut-off occurs when the ground state of the dot is no longer a bound state.}
\label{newforster}
\end{figure}

\begin{figure}[h]
\caption{(a) Energy splitting $\Delta_0$ of the qubit exciton states $\ket{\Psi_{01}}$, and $\ket{\Psi_{10}}$  in the
absence of the F\"orster interaction as a function of  the  different dots sizes $a$ and $b$.
The splitting is independent of interdot distance.
(b) The size of the mixing component of the wavefunction, $c_1$, as a function of the dot size ratio. $c_1$ has been
scaled by its dependence on the interdot distance, $R$, and typical atomic spacing, $x$.}
\label{Delta}
\end{figure}


\begin{thebibliography}{19}
\expandafter\ifx\csname natexlab\endcsname\relax\def\natexlab#1{#1}\fi
\expandafter\ifx\csname bibnamefont\endcsname\relax
  \def\bibnamefont#1{#1}\fi
\expandafter\ifx\csname bibfnamefont\endcsname\relax
  \def\bibfnamefont#1{#1}\fi
\expandafter\ifx\csname citenamefont\endcsname\relax
  \def\citenamefont#1{#1}\fi
\expandafter\ifx\csname url\endcsname\relax
  \def\url#1{\texttt{#1}}\fi
\expandafter\ifx\csname urlprefix\endcsname\relax\def\urlprefix{URL }\fi
\providecommand{\bibinfo}[2]{#2}
\providecommand{\eprint}[2][]{\url{#2}}

\bibitem[{\citenamefont{F{\"{o}}rster}(1959)}]{forster59}
\bibinfo{author}{\bibfnamefont{T.}~\bibnamefont{F{\"{o}}rster}},
  \bibinfo{journal}{Disc. Farad. Soc.} \textbf{\bibinfo{volume}{27}},
  \bibinfo{pages}{7} (\bibinfo{year}{1959}).

\bibitem[{\citenamefont{Dexter}(1953)}]{dexter53}
\bibinfo{author}{\bibfnamefont{D.~L.} \bibnamefont{Dexter}},
  \bibinfo{journal}{J.~Chem.~Phys.} \textbf{\bibinfo{volume}{21}},
  \bibinfo{pages}{836} (\bibinfo{year}{1953}).

\bibitem[{\citenamefont{Hu et~al.}(2002)}]{hu02}
\bibinfo{author}{\bibfnamefont{X.}~\bibnamefont{Hu}} \bibnamefont{et~al.},
  \bibinfo{journal}{Quarterly Rev. of Biophysics}
  \textbf{\bibinfo{volume}{35}}, \bibinfo{pages}{1} (\bibinfo{year}{2002}).

\bibitem[{\citenamefont{Crooker et~al.}(2002)}]{crooker02}
\bibinfo{author}{\bibfnamefont{S.~A.} \bibnamefont{Crooker}}
  \bibnamefont{et~al.}, \bibinfo{journal}{Phys. Rev. Lett.}
  \textbf{\bibinfo{volume}{89}}, \bibinfo{pages}{186802}
  (\bibinfo{year}{2002}).

\bibitem[{\citenamefont{Hettich et~al.}(2002)}]{hettich02}
\bibinfo{author}{\bibfnamefont{C.}~\bibnamefont{Hettich}} \bibnamefont{et~al.},
  \bibinfo{journal}{Science} \textbf{\bibinfo{volume}{298}},
  \bibinfo{pages}{385} (\bibinfo{year}{2002}).

\bibitem[{\citenamefont{Guest et~al.}(2001)}]{gammon01}
\bibinfo{author}{\bibfnamefont{J.~R.} \bibnamefont{Guest}}
  \bibnamefont{et~al.}, \bibinfo{journal}{Science}
  \textbf{\bibinfo{volume}{293}}, \bibinfo{pages}{2224} (\bibinfo{year}{2001}).

\bibitem[{\citenamefont{Lovett et~al.}()}]{lovett02}
\bibinfo{author}{\bibfnamefont{B.~W.} \bibnamefont{Lovett}}
  \bibnamefont{et~al.}, \eprint{in preparation}.

\bibitem[{\citenamefont{Kamada et~al.}(2001)}]{kama01}
\bibinfo{author}{\bibfnamefont{H.}~\bibnamefont{Kamada}} \bibnamefont{et~al.},
  \bibinfo{journal}{Phys. Rev. Lett.} \textbf{\bibinfo{volume}{87}},
  \bibinfo{pages}{246401} (\bibinfo{year}{2001}).

\bibitem[{\citenamefont{Biolatti et~al.}(2002)}]{biolatti02}
\bibinfo{author}{\bibfnamefont{E.}~\bibnamefont{Biolatti}}
  \bibnamefont{et~al.}, \bibinfo{journal}{Phys.~Rev.~B}
  \textbf{\bibinfo{volume}{65}}, \bibinfo{pages}{075306}
  (\bibinfo{year}{2002}).

\bibitem[{\citenamefont{Harrison}()}]{harrison00}
\bibinfo{author}{\bibfnamefont{P.}~\bibnamefont{Harrison}},
  \eprint{\emph{Quantum Wells, Wires and Dots} (Wiley, New York, 2001)}.

\bibitem[{\citenamefont{Pryor et~al.}(1997)}]{pryor97}
\bibinfo{author}{\bibfnamefont{C.}~\bibnamefont{Pryor}} \bibnamefont{et~al.},
  \bibinfo{journal}{Phys. Rev. B} \textbf{\bibinfo{volume}{56}},
  \bibinfo{pages}{10404} (\bibinfo{year}{1997}).

\bibitem[{\citenamefont{Rinaldis et~al.}(2002)}]{derinaldis02}
\bibinfo{author}{\bibfnamefont{S.~D.} \bibnamefont{Rinaldis}}
  \bibnamefont{et~al.}, \bibinfo{journal}{Phys.~Rev. B}
  \textbf{\bibinfo{volume}{65}}, \bibinfo{pages}{081309}
  (\bibinfo{year}{2002}).

\bibitem[{\citenamefont{Borri et~al.}()}]{nsdeco}
\bibinfo{author}{\bibfnamefont{P.}~\bibnamefont{Borri}} \bibnamefont{et~al.},
  \eprint{Phys.~Rev.~Lett. {\bf 87}, 157401 (2001); M. Bayer and A. Forchel,
  Phys.~Rev.~B. {\bf 65}, 41308(R) (2002)}.

\bibitem[{NIS()}]{NISTsite}
\eprint{http://physics.nist.gov/Pubs/AtSpec/node17.html}.

\bibitem[{\citenamefont{Roos et~al.}(1999)}]{blatt99}
\bibinfo{author}{\bibfnamefont{C.}~\bibnamefont{Roos}} \bibnamefont{et~al.},
  \bibinfo{journal}{Phys. Rev. Lett.} \textbf{\bibinfo{volume}{83}},
  \bibinfo{pages}{4713} (\bibinfo{year}{1999}).

\bibitem[{\citenamefont{White et~al.}(2002)}]{bill02s}
\bibinfo{author}{\bibfnamefont{A.~G.} \bibnamefont{White}}
  \bibnamefont{et~al.}, \bibinfo{journal}{Phys. Rev. A}
  \textbf{\bibinfo{volume}{65}}, \bibinfo{pages}{012301}
  (\bibinfo{year}{2002}).

\bibitem[{\citenamefont{Benjamin}(2000)}]{benjamin00}
\bibinfo{author}{\bibfnamefont{S.~C.} \bibnamefont{Benjamin}},
  \bibinfo{journal}{Phys.~Rev. A} \textbf{\bibinfo{volume}{61}},
  \bibinfo{pages}{020301} (\bibinfo{year}{2000}).

\bibitem[{\citenamefont{Xie et~al.}(1995)}]{xie95}
\bibinfo{author}{\bibfnamefont{Q.}~\bibnamefont{Xie}} \bibnamefont{et~al.},
  \bibinfo{journal}{Phys.~Rev.~Lett.} \textbf{\bibinfo{volume}{75}},
  \bibinfo{pages}{2542} (\bibinfo{year}{1995}).

\bibitem[{\citenamefont{Lidar et~al.}(1998)}]{lidar98}
\bibinfo{author}{\bibfnamefont{D.~A.} \bibnamefont{Lidar}}
  \bibnamefont{et~al.}, \bibinfo{journal}{Phys. Rev. Lett.}
  \textbf{\bibinfo{volume}{81}}, \bibinfo{pages}{2594} (\bibinfo{year}{1998}).

\end{thebibliography}
\end{document}